\documentclass[a4paper]{jpconf}
\usepackage{graphicx}
\begin{document}
\title{Powering the jets in NGC\,1052}

\author{Eduardo Ros$^1$ and Matthias Kadler$^{2,3,4,5}$}

\address{$^1$ Max-Planck-Institut f\"ur Radioastronomie, Auf dem H\"ugel 69, 53121 Bonn, Germany}
\address{$^2$ Dr.\ Remeis-Sternwarte Bamberg, Friedrich-Alexander-Universit\"at Erlangen, Sternwartstrasse 7, D-96049 Bamberg, Germany} 
\address{$^3$ CRESST/NASA Goddard Flight Space Center, 662 Greenbelt, MD 20771, USA}
\address{$^4$ Universities Space Research Association, 10211 Wincopin Circle, Suite 500 Columbia, MD 20144, USA}
\address{$^5$ Erlangen Centre for Astroparticle Physics, Erwin-Rommel-Str.\ 1, D-91058 Erlangen, Germany} 

\ead{ros@mpifr.de, matthias.kadler@sternwarte.uni-erlangen.de}

\begin{abstract}
We have studied the inner regions of the LINER galaxy NGC\,1052 since the mid
1990s at high resolution with 15\,GHz very-long-baseline
interferometry observations.  A compact, two-sided jet structure is revealed,
with multiple sub-parsec scale features moving outward 
from the central region with typical speeds of 0.26\,$c$.
Complementary to this, since early 2005 we are performing 
a multi-mission campaign of observations of this source, including 
X-ray spectroscopy,
X-ray, and radio flux density monitoring, and VLBA observations
at 22\,GHz and 43\,GHz.  X-ray variability is present at time
scales of weeks, comparable with the structural changes observed
by VLBI.   Here we present first results of the high-resolution 
imaging
observations and discuss these findings in the context of
the multi-band campaign.
\end{abstract}

\section{High-resolution and the AGN engines}

The highest resolution observations in astronomy 
are possible via very-long-baseline
interferometry (VLBI), which can be conducted
routinely up to frequencies of
86\,GHz, yielding sharp images of the brightest objects in the
Universe resolved up to some tens of microarcseconds ($\mu$as; see,
e.g., the latest pioneering results by Doeleman \textit{et al} 2008 at even
higher frequencies).  Typical
resolutions are of the order of 0.5--3\,mas at the frequencies
where atmospheric and ionospheric effects do not severely affect the
performance of the observations.  High resolution observations 
of active galactic nuclei
(AGN) provide access to the regions close to the central engine, believed
to be a super-massive black hole (e.g., Zensus 1995).  
Those are complemented by single-dish flux density monitoring
to follow the changes in the innermost part also from the spectra,
which reveal variations of the  optically-thin (synchrotron) and
optically thick regions (closer to the self-absorbed jet base).
Additionally, radio spectra turnover frequency imaging yields 
information on the jet physics, particle density, composition, etc. 
(e.g., Lobanov 1998).  Polarisation observations at the higher frequencies
can also reveal the direction and nature of the magnetic fields in
the inner jet.

Following the standard model (e.g., Marscher 2006)
the emission of a radio loud AGN 
at different radio wavelengths 
originates at different distances from the 
base of the jet.
At higher frequencies, the bulk emission comes from closer 
to the central engine, since
the synchrotron-self-absorption of  the inner part of the jet is
reduced at shorter wavelengths.
Assuming a central black hole surrounded by an accretion disk, and 
perpendicularly to this, a jet with superluminal plasma accelerated, 
the radio emission is produced by synchrotron at a region where no 
self-absorption is present.  

High-resolution radio observations are complemented by other regions 
of the spectra.  AGN broadband spectra show absorption, reflection and
emission lines as well as a power-law continuum radiation which can
be studied by analyzing the spectral energy distribution.  Their spectra
show radio variability.  In the X-rays, the observations of the iron
K$\alpha$ line at 6.4\,keV (the most prominent fluorescent line) probes the
AGN circumnuclear environment, from its variability and profile.

The combination of radio and X-ray has been successful, i.e., in the
case of 3C\,120 (Marscher \textit{et al} 2002), where X-ray flux
density ``dips'' precede VLBI jet ejections.  The time delays are
of the order of weeks to months caused by the offset of the core from the
central engine and by the core shift between the observation 
frequency and the self-absorption turnover frequency.  
These 
``dips'' can be put in the context of the discoveries in 
X-ray binaries (see Fender \& Belloni 2004).  

\section{NGC\,1052: the bridge between Type 1 and Type 2 sources}

Flux-density selected samples of radio-loud AGN are usually
dominated by Type 1 sources,
with a one-sided jet morphology in high resolution radio
images and the counter-jet usually not seen due to Doppler relativistic
boosting
(see e.g., Ros, these proceedings).   Type 1
sources  have an accretion disk which is seen face-on, and no 
absorption is seen in the soft region of the X-ray spectrum. 
Type 2 sources have an
edge-on accretion disk, detected by strong absorption in the soft
X-rays (Fabian \textit{et al} 2000), and free-free 
absorption is observed at the multi-wavelength
radio structure imaging close to the central engine.

NGC\,1052 (B0238$-$084, J0241$-$0815), classified as low-ionization nuclear 
emission-line region object (LINER), is a key source between 
both types, since it is radio loud, oriented in the plane of the sky,
and emits in X-rays. 
Due  to its closeness (with a redshift of 
$z$=0.0049 (Knapp \textit{et al} 1978), NGC\,1052, despite of its 
relatively low radio luminosity, is still very bright in radio wavelengths
and can be detected at all observing VLBI bands in jet and
counter-jet (see, e.g., Lee \textit{et al} 2008 for 86\,GHz for the 
highest available VLBI frequency results published, with just one
bright component).
At arcsecond scales (Kadler \textit{et al} 2004a; 
Cooper \textit{et al} 2008) it presents an east-west faint
emission (of 30$^{\prime\prime}$ in extension) with a dominating
core, which is also seen on  mas-scales with a double-sided radio
with a position angle of $\sim$65$^\circ$.
The jet axis is considered to be nearly parallel to the plane
of the sky, with values between a lower limit of 
57$^\circ$ and 72$^\circ$ (Vermeulen \textit{et al} 2003;
Kadler \textit{et al} 2004b).
Sawada-Satoh \textit{et al} (2008) determine values around 76$^\circ$--80$^\circ$.
High opacities are measured toward the innermost component of the western
jet (Kadler \textit{et al} 2004b; Kameno \textit{et al} 2001;
Vermeulen \textit{et al} 2003), and 
also at the inner edge of the eastern jet at centimetre
wavelengths (Kadler \textit{et al} 2004b; Sawada-Satoh \textit{et al} 2008).
Millimetre observations can see through this screen and
reveal some structure at the innermost region.
Emission from H$_2$O spots has been detected both at jet and
counter-jet (Sawada-Satoh \textit{et al} 2008 and references therein), 
redshifted by 50--350\,km\,s$^{-1}$ with respect to the systemic
velocity of the galaxy.  This water emission is interpreted to be 
located at the region where opacity dominates due to 
the thermal plasma in the obscuring
torus via free-free absorption.
X-ray spectra further support the presence of a dense gas torus
(Guainazzi \& Antonelli 1999; Weaver \textit{et al} 1999; 
Kadler \textit{et al} 2004a) toward the nuclear X-ray source.

We started in 2005 a multi-mission campaign to track the ejection of
new features at the base of the jet and counter-jets and compare 
with brightness monitoring and spectral studies in X-rays and radio,
with the goal of establishing a relationship between changes in
the high energy and the jet production in AGN. 
First results of the campaign were presented in 
Kadler \textit{et al} (2006) and Ros \textit{et al} (2007).

\begin{figure}[p]
\begin{center}
\includegraphics[clip,width=0.95\textwidth]{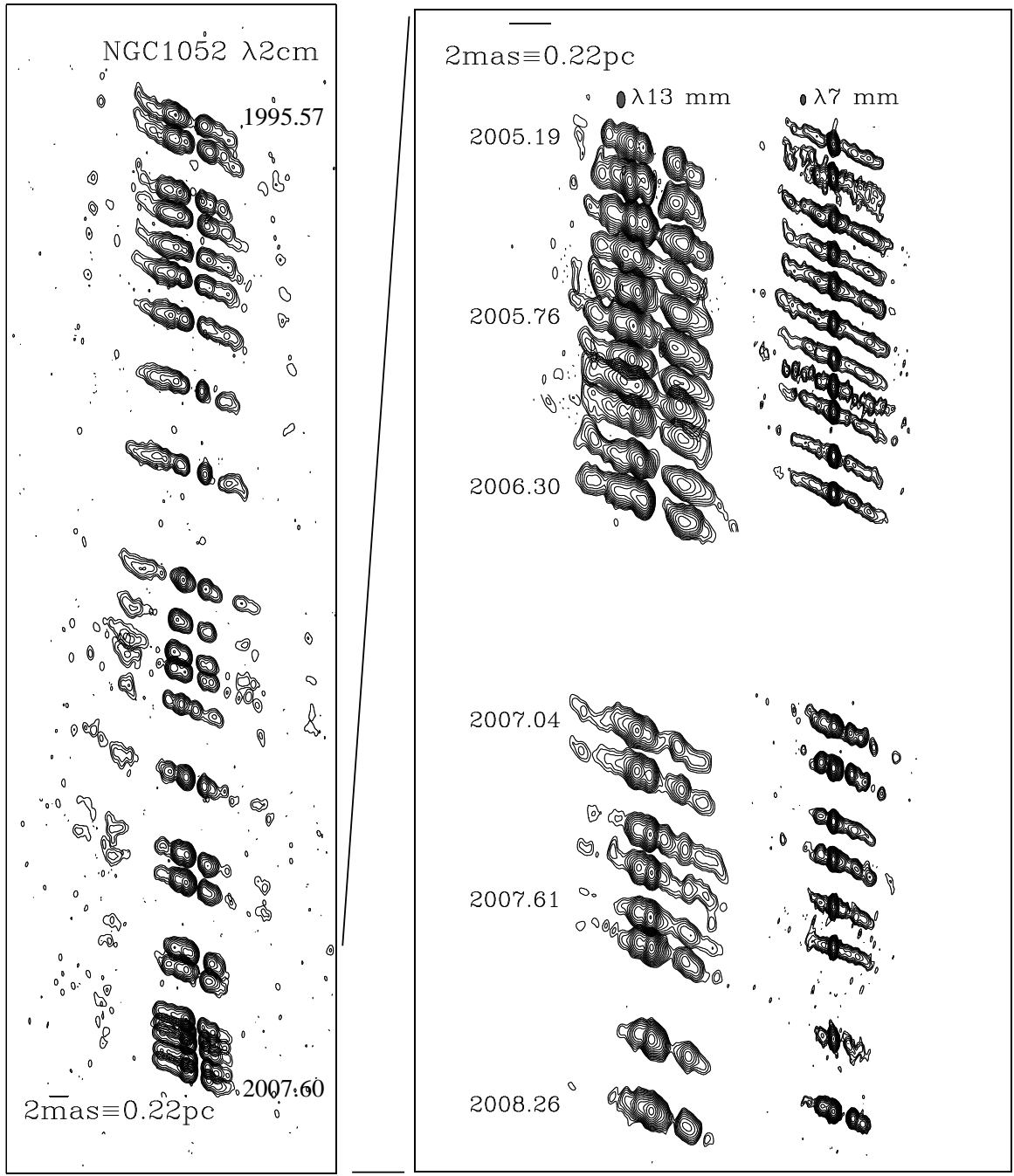}
\end{center}
\caption{\label{fig:allmaps} VLBA images of NGC\,1052.  The images
shown are spaced by their relative time intervals.  The contours
are logarithmic with steps of factor 2.  The alignment is
relative to the emission gap.  At the left panel we show images
from the 2\,cm~Survey/MOJAVE programme, registered over a decade.
At the right panel we show preliminary images from our 
monitoring programme at high frequencies, during three years 
since early 2005.  
The images have been convolved with common restoring beams, being
$1.0\times0.5$\,mas in P.A.\ for the $\lambda$2\,cm images.  The
beam for the shorter wavelengths are shown as grey ellipses next
to the wavelength label.
The lowest contours are set accordingly to the noise level individually
for each epoch, with 
values of some mJy\,beam$^{-1}$.
}
\end{figure}

\subsection{Observations}

\subsubsection{$\lambda$2\,cm VLBA observations}

\begin{figure}
\begin{center}
\includegraphics[clip,width=0.95\textwidth]{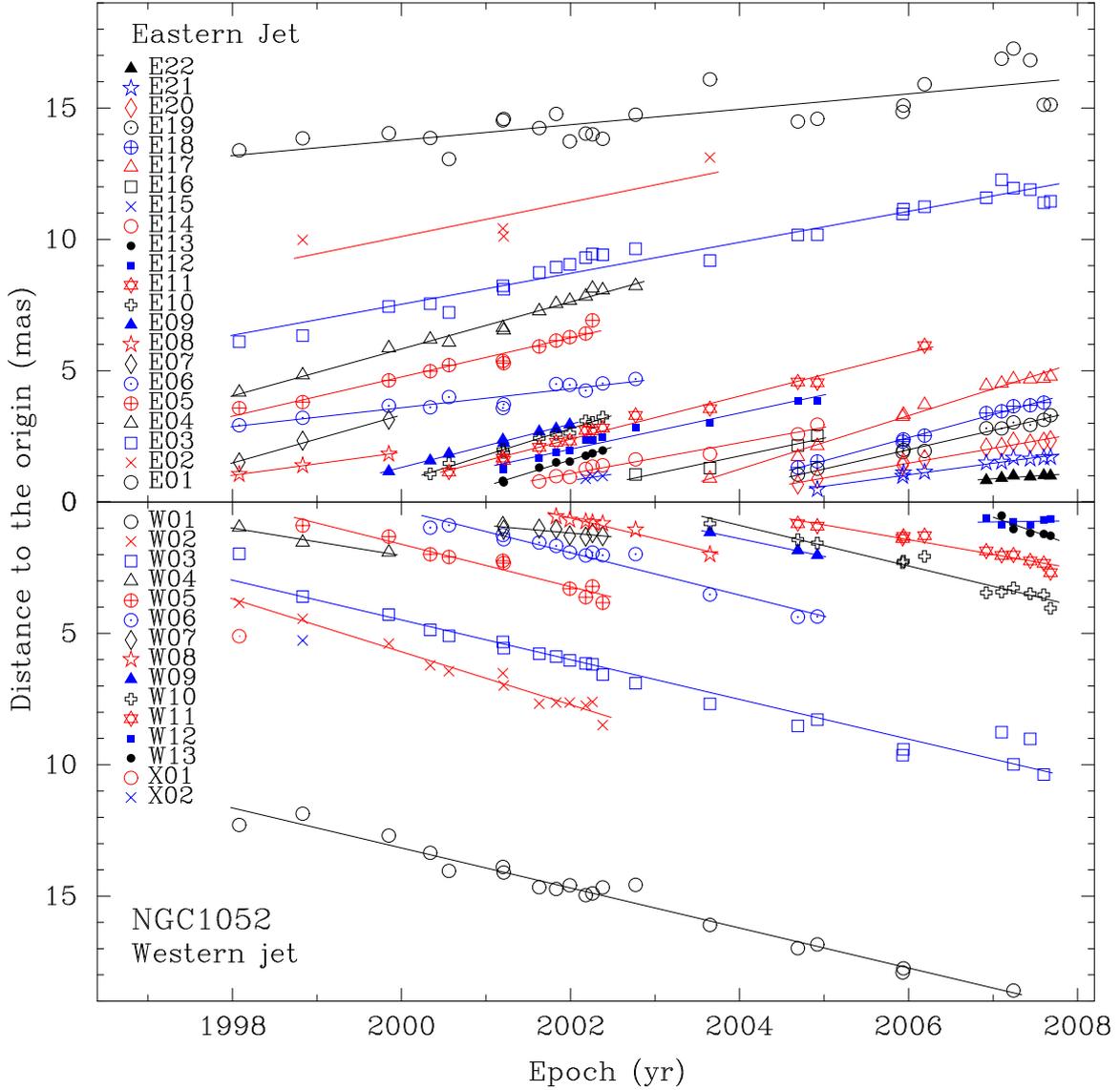}
\end{center}
\caption{\label{fig:kinematics} Model fitting results to the parsec-scale
twin-jet radio structure of NGC\,1052
at 15\,GHz.  The relative distances to the central region are
shown as a function of time for the jet (top) and the counter-jet
(bottom).  
The solid lines are linear regression fits to the data points,
and their slopes give the reported proper motions.
The typical speeds observed are of 
$\approx$0.26\,$c$ both in jet and counter-jet.  Table \ref{table:kinematics2cm} 
shows the speed values for selected components.
}
\end{figure}

NGC\,1052 has been observed with the VLBA by the 2\,cm\,Survey/MOJAVE 
programme since 1995, with typical resolutions of 1$\times$0.5\,mas
in P.A.\ 0$^\circ$.  The images resulting from these measurements
are shown in the left panel of Fig.\ \ref{fig:allmaps}.
To reproduced the structural variations in the jets of NGC\,1052,
we modelled the observed radio visibilities with Gaussian circular
functions.  We cross-identified the modelled features from epoch to epoch
by criteria of continuity in the positions and flux densities.  A
complex data base of functions results, with over 30 components 
along the more than 10 years of observations.
In Fig.\ \ref{fig:kinematics} we show the proper motions of these
model-fitted features (the observations and results 
are part of the complete kinematic analysis of the
MOJAVE/2\,cm\,Survey program (M L Lister \textit{et al}, in
preparation; Ros, these proceedings). 
The reference origin to align the images has been chosen as
the middle point between the two innermost components, although we 
note
that the opacity effects can shift this position.  This sets the question of the ejection epoch for 
the emerging components, since we can extrapolate the linear fit
to be zero, but this can be considered as a lower limit.  Notice
that when computing kinematic epoch ejections for one-sided
radio sources, the origin values are usually extrapolated to the core
position ($\tau\sim1$ surface).  We show the speeds and
maximum flux densities for selected components in 
Table~\ref{table:kinematics2cm}.  A more detailed analysis on
the ejection, possible acceleration of components, non-linear motions,
etc., is going to be presented in M Kadler \textit{et al} (in
preparation) and elsewhere.

\begin{table}[thb]
\begin{center}
\caption{\label{table:kinematics2cm} Kinematic results for selected
components of the twin jet in NGC\,1052}
\begin{tabular}{ccr@{\,$\pm$\,}lr@{\,$\pm$\,}lr@{\,$\pm$\,}l}
\br
   & No of & \multicolumn{2}{c}{~} & \multicolumn{2}{c}{~} \\
ID & Epochs & \multicolumn{2}{c}{$\mu$} & \multicolumn{2}{c}{$\beta_\mathrm{app}$} & \multicolumn{2}{c}{epoch} \\ 
   & \footnotesize{[$\mu$as\,yr$^{-1}$]} & \multicolumn{2}{c}{\footnotesize{[yr]}} \\ 
\mr
E01 & 25 &  447&174 &  0.15&0.06  &   1971&13  \\ 
E03 & 26 &  595&26  & 0.196&0.009 & 1987.4&0.7 \\ 
E11 & 14 &  824&27  & 0.272&0.009 & 1999.1&0.1 \\ 
E17 & 12 & 1038&34  &  0.34&0.01  & 2002.8&0.1 \\ 
E18 & 10 &  889&27  &  0.294&0.09 & 2003.2&0.1 \\ 
\mr
W11 & 11 &  536&46  &  0.18&0.02  & 2003.3&0.3 \\ 
W06 & 14 &  813&53  &  0.27&0.02  & 1999.6&0.2 \\ 
W03 & 23 &  757&36  &  0.25&0.01  & 1994.1&0.1 \\ 
W01 & 20 &  763&30  &  0.25&0.01  & 1982.8&0.8 \\ 
\br 
\end{tabular}
\end{center}
\end{table}

\subsubsection{$\lambda$$\lambda$1.3\,cm and 0.7\,cm VLBA Observations}
We have been monitoring 
the structure of NGC\,1052 at 22\,GHz
and 43\,GHz using the VLBA every six weeks (with some larger gaps
in the time sampling due to scheduling reasons) since early 2005.  The observations at
these frequencies have different dynamic ranges and beams due to the
different constraints during the observing epochs (e.g., weather constraints
or failure of one or two stations of the VLBA).  We present contour
plots of the preliminary images convolved with a common beam in the 
right panel of Fig.\ \ref{fig:allmaps}.  A detailed model fitting of
the observed features is in progress, as well as the astrometric registration
of the images, since they were observed in phase-referencing mode, 
including scans on the calibrator J0243$-$0550.

The imaging results show the free-free absorption screen to be still
present at 22\,GHz and getting transparent at 43\,GHz.  New 
components can be traced when being ejected from the innermost
part of the jet.  The light curves of the single-dish telescope measurements
show a decay from 2005 to early 2008, with some hints of a rising flux
density for the latest months (H Aller and M Aller, private communication). 

\section{Results}

Our long-term 15\,GHz results confirm the morphology and kinematics
reported in Vermeulen \textit{et al} (2003), with prominent 
jet and counter-jet features moving downstream with mildly relativistic
speeds of $\approx$0.25\,$c$.  The source has experienced a decrease
in flux density at all frequencies since 2005.  This is caused by
the fading out of the features travelling outward the jet, which is
not compensated by new, strong components at the base of the jet.
There are some hints that this situation is beginning to change from
the early 2008 observations.

The results of the multi-band campaign, including \textit{RXTE} light
curves and spectra, \textit{XMM-Newton}, \textit{Suzaku}, and 
\textit{Chandra} observations, will be published elsewhere.  
Our imaging campaign at the highest frequencies is planned to last at 
least until early 2009.  The MOJAVE program will be also continued 
beyond that and at least until 2010, providing observations 
of NGC\,1052 at 15\,GHz every several months.

\ack
The VLBA is operated by the National Radio Astronomy Observatory, which 
is a facility of the USA National Science Foundation operated under 
cooperative agreement with Associated Universities, Inc.

\section*{References}
\begin{thereferences}
\item Cooper N J, Lister M L and Kochanczyk M D 2008 \textit{ApJS} \textbf{171} 376--388
\item Doeleman S D \textit{et al} 2008 \textit{Nature} \textbf{455} 78--80
\item Fabian A C, Iwasawa K, Reynolds C S and Young A 2000 \textit{PASP} \textbf{112} 1145--1161
\item Fender R and Belloni T 2004 \textit{ARA\&A} \textbf{42} 317
\item Guainazzi M A and Antonelli L A 1999 \textit{MNRAS} \textbf{304} L15--L19 
\item Kadler M, Kerp J, Ros E, Falcke H, Pogge R W and Zensus J A 2004a \textit{A\&A} \textbf{420} 467--  
\item Kadler M, Ros E, Lobanov A P, Falcke H and Zensus J A 2004b \textit{A\&A} \textbf{426} 481--493 
\item Kadler M \textit{et al} 2006 \textit{Challenges of Relativistic Jets} \texttt{http://www.oa.uj.edu.pl/2006jets/posters/Kadler/Kadler.pdf}
\item Kameno S, Sawada-Satoh S, Inoue M, Shen Z Q and Wajima K \textit{PASJ} \textbf{53}  169--178
\item Lee S S, Lobanov A P, Krichbaum T P, Witzel A, Zensus J A, Bremer M, Greve A and Grewing M 2008 \textit{AJ} \textbf{136} 159--180 
\item Lobanov A P \textit{A\&A} \textbf{330} 79--89
\item Marscher A P, Jorstad S G, G\'omez J L, Aller M F, Ter\"astranta H, Lister M L and Stirling A M 2002 \textit{Nature} \textbf{417} 625--
\item Marscher A P 2006 \textit{Relativistic Jets} ed P A Hughes and J N Bregman, (New York: American Institute of Physics) pp 1--22
\item Ros E, Kadler M, Kaufmann S, Kovalev Y Y, Tueller J and Weaver K A 2007 \textit{Highlights of Spanish Astrophysics IV} ed F Figueras \textit{et al} (Dordrecht: Springer) p 165
\item Sawada-Satoh S, Kameno S, Nakamura K, Namikawa D, Shibata K M and Inoue M 2008 \textit{ApJ} \textbf{680} 191--199 
\item Vermeulen R C, Ros E, Kellermann K I, Cohen M H, Zensus J A and van Langevelde H J 2003 \textit{A\&A} \textbf{401} 113-127 
\item Weaver K A, Wilson A S, Henkel C and Braatz J A 1999 \textit{ApJ} \textbf{520} 130--
\item Zensus J A 1995 \textit{AR\&A} \textbf{35} 607
\end{thereferences}

\end{document}